\documentclass[
  prd,aps,showpacs,groupedaddress,eqsecnum,notitlepage,nofootinbib,superscriptaddress,
  colorlinks=true, linkcolor=blue, citecolor=blue, urlcolor=blue, pdfstartview=FitV, breaklinks=true
]{revtex4-2}

\usepackage[utf8]{inputenc}
\usepackage[T1]{fontenc}
\usepackage{lmodern}

\usepackage{amsmath,amssymb,amsfonts,bm,tensor,slashed}

\usepackage{graphicx}
\usepackage{float}
\usepackage{epstopdf}
\usepackage{subfigure}
\usepackage{tikz}

\usepackage{color}
\usepackage{enumerate}
\usepackage{siunitx}
\usepackage{scalerel}
\usepackage{accents}
\usepackage{orcidlink}

\usepackage{hyperref}

\begin{document}

\title{New Solutions for Topological Defects with Continuous Distributions:\\ A Conformal Metric Perspective}

\author{A.~M.~de~M.~Carvalho\,\orcidlink{0009-0006-3540-0364}}
\email{alexandre@fis.ufal.br}
\affiliation{Instituto de F\'{\i}sica, Universidade Federal de Alagoas, 57072-970 Macei\'o, AL, Brazil}

\author{G.~Q.~Garcia\,\orcidlink{0000-0003-3562-0317}}
\email{gqgarcia99@gmail.com}
\affiliation{Centro de Ci\^encias, Tecnologia e Sa\'ude, Universidade Estadual da Para\'iba, 58233-000 Araruna, PB, Brazil}

\author{C.~Furtado\,\orcidlink{0000-0002-3455-4285}}
\email{furtado@fisica.ufpb.br}
\affiliation{Departamento de F\'isica, Universidade Federal da Para\'iba, 58051-970 Jo\~ao Pessoa, PB, Brazil}


\begin{abstract}
We present new exact solutions for two-dimensional geometries generated by continuous distributions of topological defects within a conformal metric framework. By reformulating Einstein’s equations in two dimensions as a Poisson equation for the conformal factor, we analyze how smooth defect densities—such as Gaussian, exponential, and power-law profiles—regularize curvature singularities and encode nontrivial topological information. Each distribution yields a well-defined geometry that interpolates between localized curvature near the defect core and asymptotic flatness. We compute the Ricci scalar and total curvature, confirming consistency with the Gauss-Bonnet theorem. Our results provide a unified geometric description of regularized disclination-like defects and offer insights into analog gravity, crystalline materials, and two-dimensional systems with emergent curvature.
\end{abstract}
\maketitle
\section{Introduction}

Topological defects represent singularities in ordered systems, emerging in contexts where topological constraints hinder the continuous maintenance of this state. In materials such as liquid crystals, these defects correspond to regions where molecular alignment is disrupted, influencing the optical, elastic, and dynamic properties of the system~\cite{Fumeron2021, Fumeron2023b, mermin1979topological}. Traditionally, the theoretical description of these defects relies on delta distributions, modeling them as point-like concentrations of curvature and stress. However, this approach introduces discontinuities and mathematical singularities in energy and curvature, limiting its applicability to real physical systems. This issue has been studied in general relativity, where distributional sources are known to generate ill-defined geometries~\cite{geroch1987strings,steinbauer2006use}. A more realistic and robust alternative is to consider continuous defect distributions, allowing for a gradual transition in spatial curvature and avoiding problematic issues such as divergences in energy and momentum. These distributions better reflect how defects arise in experiments, where factors such as preparation conditions or interactions with external fields lead to finite spatial distributions rather than singular points~\cite{seung1988defects,bowick2009two}.

Choosing continuous distributions for defects that naturally regularize the geometry ensures that the conformal factor satisfies a well-behaved Poisson equation without creating artificial discontinuities. Additionally, this approach enables direct connections with geometric models in the Katanaev-Volovich formulation~\cite{KatanaevVolovich1992, katanaev2005geometric}, where topological defects are interpreted as curvature sources similar to masses in (2+1)-dimensional gravity theories. This perspective aligns with the foundational work of Deser and Jackiw~\cite{DeserJackiw1984, DeserJackiw1989}, where point-like defects in three-dimensional gravity were studied as sources of curvature within a constant curvature background. Their research on conical singularities and localized string-like sources provides a basis for understanding how defects interact with the surrounding geometry. Although their approach relies on delta function sources that cause singularities, we extend this framework by considering smooth defect distributions, leading to well-behaved curvature that avoids singularities. Similar methods have been suggested in lower-dimensional gravity~\cite{deser1984three,carlip1998quantum} and in elastic theory using geometric gauge formulations~\cite{kleinert1989gauge}.

Reformulating Einstein’s equation in terms of the conformal factor allows us a direct analogy with solutions in (2+1)-dimensions~\cite{geroch1987strings,deser1984three}. These studies also highlight the challenges of handling distributional sources in general relativity and motivate the use of regularized geometries. Beyond being an elegant mathematical approach, this modeling allows for an accurate description of physical systems exhibiting continuous spatial deformations, such as elastic materials, liquid crystals, and two-dimensional surfaces under mechanical stresses~\cite{nelson2002defects,zurek1996cosmological}. Similar methods have been successfully used in graphene physics, where lattice deformations produce effective gauge fields and curvature effects that are described within a geometric and topological framework~\cite{Cortijo2007}. By transitioning from localized defects to continuous distributions, we aim to bridge the gap between the traditional conical defect models with more realistic cases where defects span a finite spatial extension. 

Our research demonstrates how smooth defect densities influence the curvature and topology of two-dimensional systems, their connection to gravitational analogies, and the modeling of topological defects in two-dimensional space using a conformal metric, analyzing continuous defect density distributions. Specifically, we examine two distinct configurations: a Gaussian distribution, where defect density concentrates in a central region with a smooth decay toward the edges, and an exponential distribution, which features a gradual decrease in defect density as the distance from the defect center increases. This article is organized as follows. In Section II, we present the theoretical framework of topological defects in two-dimensional conformal geometries, highlighting the relationship between defect density and spatial curvature~\cite{Eguchi1980,nakahara2003geometry}. In Section III, we solve the Poisson equation for Gaussian and exponential distributions, detailing the behavior of the conformal factor and related geometric properties. In Section IV, we discuss the implications of these solutions, exploring the impact of using a continuous distribution to describe the curvature induced by defects. Finally, in Section V, we summarize our conclusions and suggest possible directions for future applications and extensions of the model. In particular, we emphasize that the curvature integrals are consistent with the Gauss-Bonnet theorem~\cite{chern1944gauss}, which guarantees topological coherence even in the presence of smooth and non-quantized curvature distributions.

\section{Geometric Theory of Defects}

Using a conformal metric to analyze topological defects like dislocations and disclinations is a natural approach in both continuum mechanics and emergent gravity theories, where local geometric changes can be viewed as sources of curvature. Specifically, in (2+1)-dimensional gravity, these defects are similar to masses that cause distortions in space, creating an effective gravitational effect. Just as a mass produces curvature in traditional gravitational theories, topological defects alter the surrounding space, resulting in a geometric field that can be described using a conformal metric, which makes the mathematical analysis of these deformations~\cite{KatanaevVolovich1992,katanaev2005geometric,kleinert1989gauge}. The conformal metric can be expressed as
\begin{eqnarray}
    ds^2 = e^{2\Omega(r,\theta)}(dr^2 + r^2 d\theta^2) + dz^2,
    \label{metricaconf}
\end{eqnarray}
where \(\Omega(r, \theta)\) is the conformal factor. Here, Katanaev and Volovich replace the temporal variable \( t \) with the spatial coordinate \( z \) because their geometric theory of defects models static deformations in three-dimensional solids. The \( z \)-axis represents the direction along the defect, such as in screw dislocations, making \( z \) analogous to time in (2+1)-dimensional gravity. This substitution simplifies the mathematical framework by treating defects as sources of curvature and torsion in a static spatial setting, without requiring a dynamic description of space-time~\cite{hehl1976spin}. The metric must satisfy the Einstein equation, expressed as
\begin{eqnarray}
    R_{\alpha\beta} - \frac{1}{2}g_{\alpha\beta}R = -8\pi G T_{\alpha\beta},
\end{eqnarray}
where $R_{\alpha\beta}$ is the Ricci tensor, $R$ is the Ricci scalar and $G$ is the gravitational constant. Moreover, $G$ in condensed matter is a constant associated with the continuum elastic medium, $T_{\alpha \beta}$ is the source of stress and strain fields, {\it i.e.}, the density of defects itself. And finally, we see that $g_{\alpha\beta}$ is the metric tensor.

We introduce an appropriate dual $1$-form basis (coframe), defined in terms of local \textit{zweibein} fields by $\hat{e}^{a} = e^{a}_{\ \mu}(x) dx^{\mu}$~\cite{Eguchi1980}, where
\begin{subequations}
\begin{eqnarray}
\hat{e}^{1}&=& \exp(\Omega)\,dr \mbox{,}  \label{e1}\\
\hat{e}^{2}&=& \exp(\Omega)\,r\,d\theta \mbox{.}  \label{e2}
\end{eqnarray}
\end{subequations}
The metric (\ref{metricaconf}) can be decomposed into \textit{zweibein} fields as $g_{\mu \nu }=\delta_{a b}\,e^{a}_{\ \mu}\,e^{b}_{\ \nu}$, so that the geometry is locally flat. The greek indices will be raised and lowered with $g^{\mu\nu}$ and $g_{\mu\nu}$ and the latin ones will be raised and lowered by $\delta^{ab}$ and $\delta_{ab}$. The tetrad field and its inverse for the 1-form basis, which corresponds to the metric (\ref{metricaconf}), can be written as the following matrices:
\begin{eqnarray}
E^{a}_{\ \mu}=\mbox{diag}\left(e^{\Omega}, e^{\Omega}\,r\right)\ \ \text{and}\ \ E_{a}^{\ \mu} =\mbox{diag}\left(e^{-\Omega}, e^{-\Omega}/r\right).
\end{eqnarray}
We introduce an affine spin connection 1-form $\omega^{a}_{\ b}$ and define the torsion 2-form and the curvature 2-form, respectively, by
\begin{subequations}
\begin{eqnarray}
T^{a}&=&\frac{1}{2}T^{a}_{\ bc}\;\hat{e}^{b}\wedge \hat{e}^{c}=d\hat{e}^{a}+\omega^{a}_{b} \wedge \hat{e}^{b}\mbox{,} \\
R^{a}_{\ b}&=&\frac{1}{2}R^{a}_{\ bcd}\; \hat{e}^{c}\wedge \hat{e}^{d}.
\end{eqnarray}
\end{subequations}
These expressions are known as the Maurer-Cartan structure equations. Since we already know that the geometry is Riemannian, we make use of the torsion-free condition $d\hat{e}^{a}+\omega^{a}_{\ b} \wedge \hat{e}^{b}=0$. From the 1-form basis, we can determine the non-null connection 1-forms:
\begin{eqnarray}
\label{1form}
\omega^{1}_{\ 2}&=&- \frac{1}{r}\frac{\partial \Omega}{\partial \theta}dr-\left( \frac{\partial \Omega}{\partial r}r+1 \right)d\theta  \mbox{.}
\end{eqnarray}
The connection 1-form obeys the metricity condition $\omega^{1}_{\ 2}=-\omega^{2}_{\ 1}$. The connection 1-forms transform in the same way as the gauge potential of a non-Abelian gauge theory~\cite{kleinert1989gauge}, which means that two elements of the group do not commute. The 2-form curvature can be obtained from the second Maurer-Cartan equation~\cite{Carvalho2013}
\begin{equation}
R^{1}_{\ 2}= \left( \frac{1}{r}\frac{\partial^{2}\Omega}{\partial \theta^{2}}-\frac{\partial^{2}\Omega}{\partial r^{2}}-\frac{\partial\Omega}{\partial r}\right)dr\wedge d\theta.
\end{equation}
With knowledge of the metric we can easily obtain the Ricci scalar,
\begin{equation}
R=2e^{2\Omega }\Delta \Omega.  
\label{ricci}
\end{equation}
And then we obtain the differential equation for the conformal factor,
\begin{equation}
\Delta \Omega =-\lambda , 
\label{poisson}
\end{equation}
where $\lambda$ is a parameter associated with the density of defects by $\lambda=8\pi GT_{zz}$. The conformal metric reformulates Einstein's highly nonlinear equations into a Poisson equation, which is linear and well-established in various areas of physics, facilitating analytical and numerical solutions.

The interpretation of the conformal factor $\Omega$ as a geometric potential is consistent with previous studies of topological defects on curved surfaces. In particular, Vitelli and Turner~\cite{Vitelli2004}showed that in systems such as superfluids, superconducting films, and liquid crystals deposited on curved substrates, each defect interacts with the underlying geometry via a potential governed by a Poisson equation with Gaussian curvature as its source. This structure mirrors the approach adopted here, where regularized curvature arises from smooth defect distributions instead of point-like singularities. Identifying $\Omega$ with such a geometric potential reinforces the connection between curvature, topology, and emergent forces. This suggests that our geometric regularization may be interpreted as a classical analogue of the conformal anomaly discussed in~\cite{Vitelli2004}. Moreover, adopting a continuous description of disclinations, central to our construction, finds support in recent approaches that approximate densely packed defects by a smooth distribution of curvature. In particular, Schmitz {\it et al.}~\cite{Schmitz2021} argue that when the defect density is sufficiently high and individual cores overlap, the elastic medium can be coarse-grained such that the discrete nature of the defects becomes irrelevant at larger scales. This justifies using a continuous curvature source in the Poisson equation and underlines the physical consistency of our conformal factor as a geometric potential induced by a regularized distribution of disclinations.

\section{Solution of the Poisson Equation with an Exponentially Decreasing Source Term}

The exponential distribution of defects avoids singularities, ensures a smooth curvature transition, and better reflects real physical systems. Additionally, it allows for exact analytical solutions, facilitating modeling and connections with emerging gravitational theories. We consider an exponentially decreasing distribution for \(\lambda(r)\) of the form:
\begin{eqnarray}
\lambda(r) &=& \lambda_0 e^{-r / r_0},
\label{exponencial}
\end{eqnarray}
here we have that the parameter \(\lambda_0\) is the initial defect density and \(r_0\) is the characteristic decay length.
The solution of the Poisson equation for a defect distribution that decays exponentially (\ref{exponencial}) is given by
\begin{equation}
\Omega(r) = \lambda_0 r_0^2 \text{Ei}(-r/r_0) - \lambda_0 r_0^2 e^{-r/r_0},
\label{omegagauss}
\end{equation}
where $Ei(x)$ is the exponential integral function~\cite{abramowitz1972handbook}.
At large distances, the conformal metric (\ref{metricaconf}) can be approached by the next expression:
\begin{equation}
 ds^2 = e^{-2\lambda_0 r_0^2 e^{-r/r_0}} \left( dr^2 + r^2 d\theta^2 \right).
\end{equation}

\begin{figure}[H]
    \centering
    \includegraphics[width=0.6\textwidth]{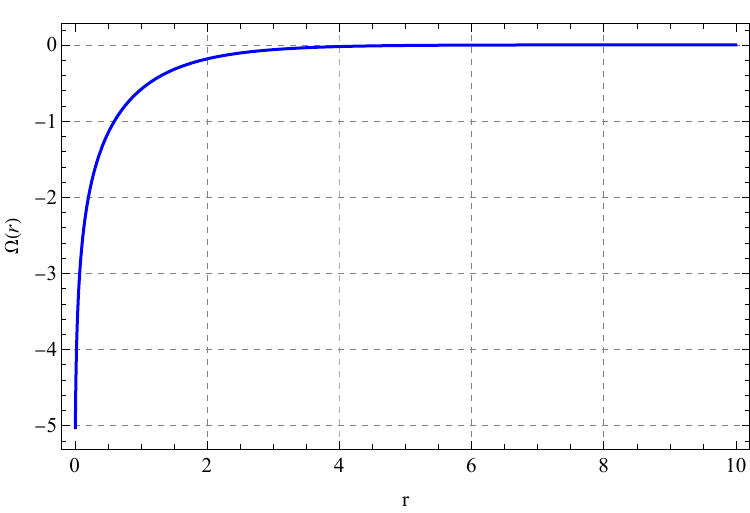}
    \caption{Conformal factor $\Omega(r)$ associated with an exponential distribution of topological defects. The expression used is $\Omega(r) = \lambda_0 r_0^2\, \mathrm{Ei}(-r/r_0) - \lambda_0 r_0^2\, e^{-r/r_0}$, where $\mathrm{Ei}(-r/r_0)$ denotes the exponential integral function.}
    \label{fig:omega_exponential}
\end{figure}

Unlike the disclination case~\cite{KatanaevVolovich1992}, where the metric is globally altered by a topological defect, the deformation caused by the exponentially decaying source remains confined to a finite region. The factor $e^{-2\lambda_0 r_0^2 e^{-r/r_0}}$ ensures that the metric modification decreases exponentially with distance. This localized effect contrasts with the global conical geometry seen in disclination-induced spaces, where an angular deficit persists at all scales. As a result, the current solution represents a localized geometric disturbance rather than a fundamental topological change. In a previous study, Culetu~\cite{culetu2022modifiedrindler} proposed a modified Rindler metric with an exponential factor to smooth out the horizon, removing singularities and ensuring smooth behavior at small distances. This modification localizes curvature effects, causing them to decay exponentially rather than persisting globally, while still recovering the standard Rindler metric at large distances. The model keeps curvature invariants finite, preventing divergences typically linked to classical Rindler space. Additionally, introducing a fundamental length scale suggests a Planck-scale modification of acceleration-induced horizons, offering a potential connection between Rindler geometry and quantum gravity effects. The figure~\ref{fig:omega_exponential} illustrates the behavior of the conformal factor $\Omega(r)$ for the exponential defect distribution. The function starts from large negative values near the core when $r \to 0$, increases monotonically, and saturates at zero from below as $r \to \infty$. This ensures that the conformal metric smoothly approaches the Euclidean metric at large distances, satisfying the asymptotic flatness condition. The regularity of $\Omega(r)$ across the entire domain contrasts with the logarithmic divergence of isolated disclinations and highlights the role of the exponential distribution in providing a physically well-behaved, globally consistent geometry.

\subsection{Topological analysis for an exponentially decreasing defect distribution}

\begin{figure}[H]
    \centering
    \includegraphics[width=0.6\textwidth]{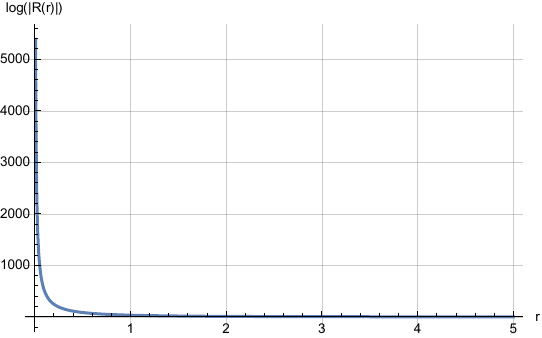}
    \caption{Logarithmic plot of the absolute value of the Ricci scalar $|R(r)|$ for the exponential distribution of topological defects. The curvature is sharply peaked near the origin and decays rapidly as $r$ increases, reflecting the compact support of the geometric deformation.}
    \label{fig:ricci_exp_log}
\end{figure}

In this subsection, we derive the expression for the Ricci scalar curvature in the context of the conformal metric employed in the study of topological defects. To ensure the consistency of our results, we apply the Gauss-Bonnet theorem, verifying that the curvature distributions adhere to the expected topological constraints. By substituting \eqref{omegagauss} into the expression for the Ricci scalar \eqref{ricci}, we obtain its asymptotic form at large distances
\begin{align}
R \approx\; & 2\lambda_0 \frac{-r^3 + 2r^2r_0 + r r_0^2 + r_0^3}{r^3} \times \nonumber \\
& \times \exp\left[-2\lambda_0 r_0^2 e^{-r/r_0} + \frac{2\lambda_0 r_0^3 e^{-r/r_0}}{r} - \frac{r}{r_0} \right].
\end{align}
We observe that the Ricci scalar exhibits an exponential decay due to the factor, ensuring that curvature effects remain localized. The polynomial term moderates this decay, creating a smooth transition between the defect region and the asymptotically Euclidean space. The figure~\ref{fig:ricci_exp_log} displays the logarithmic profile of the Ricci scalar $R(r)$ for an exponential distribution. The peak at the origin emphasizes the strong localization of the curvature near the defect core, while the exponential decay confirms that the deformation becomes negligible far from the center. This behavior aligns with the expected compactness of the Gaussian geometry and ensures that the total integrated curvature remains finite, maintaining topological neutrality at infinity.
The Gauss-Bonnet theorem for a surface without boundaries states that~\cite{eguchi1980gravitation,nakahara2003geometry}:
\begin{eqnarray}
\int_S K dA &=& 2\pi \chi(S),
\label{Gauss-Bonet}
\end{eqnarray}
where $K$ is the Gaussian curvature and $\chi(s)$ is the Euler characteristic of the surface. For this case, the Gaussian curvature can be written as 
\begin{eqnarray}
K &=& -\lambda_0 r_0 e^{-r/r_0} - \frac{1}{r}.
\end{eqnarray}
The Ricci scalar and Gaussian curvature are related through the conformal factor, as in two-dimensional spaces, they satisfy the following relation:
\begin{eqnarray}
R &=& 2K e^{2\Omega},
\end{eqnarray}
and this relation ensures that the geometric properties derived from both curvature measures are consistent within the conformal framework. The integral over the surface results in
\begin{eqnarray}
\int_S K dA &=& 2\pi \lambda_0 r_0^3 \left(\frac{1}{25} + \frac{\pi \lambda_0^2 r_0^6}{2} \right).
\end{eqnarray}
As a consequence, the above expression confirms that the topology is preserved and that the calculation of the topological charge is finite.

The review of this section confirmed the consistency of the Ricci scalar expression and the application of the Gauss-Bonnet theorem, ensuring that the regularization of the singularity at the defect center is correctly implemented. Additionally, the analysis revealed that the curvature flux can be interpreted similarly to the Aharonov-Bohm effect, reinforcing the connection between topological defects and geometry in a manner similar to the magnetic flux in quantum mechanics. In an exponentially decaying defect distribution, the defect density smoothly decreases away from the core, resulting in a curvature profile that remains finite and avoids singularities. The resulting topological charge is continuous rather than quantized, distinguishing it from isolated point defects. The use of the Gauss-Bonnet theorem ensures the consistency of this outcome, ensuring that the total curvature flux is well-defined. This kind of defect distribution is especially relevant in physical systems where topological defect effects are spread over a finite region rather than concentrated at a single point, such as in condensed matter physics and soft materials.

\section{Solution of the Poisson Equation for a Gaussian Distribution of Defects}

A key aspect of modeling topological defects in two-dimensional spaces is using smooth defect distributions to regularize singularities, similar to statistical studies of gravitational force distributions by Chavanis~\cite{Chavanis2009}. Both methods rely on the Poisson equation, with Gaussian distributions serving as a smoothing mechanism. In our work, Gaussian defect densities create a smooth curvature transition, while Chavanis finds that in one-dimensional gravity, the force distribution is purely Gaussian, and in two dimensions, it maintains a Gaussian core with power-law tails. Both studies demonstrate how continuous distributions eliminate singularities. Our model smooths curvature discontinuities, and Chavanis' work shows that averaging over mass sources results in a Gaussian-like force distribution. These parallels emphasize a deeper connection between stochastic gravitational techniques and geometric defect modeling, offering insights into how smooth distributions help regularize singularities. The defect density function can be expressed in the form of a Gaussian distribution, as in
\begin{eqnarray}
    \lambda(r) = \frac{\lambda_0}{\sqrt{\pi} r_0} e^{-r^2 / r_0^2},
\end{eqnarray}
where \( \lambda_0 \) denotes the total defect density and \( r_0 \) characterizes the decay length. To ensure physical consistency of the solution, it is important to emphasize that the metric remains regular at the origin and asymptotically recovers the Euclidean metric at large distances \( r \to \infty \). This prevents artificial singularities and guarantees that the solution is well-behaved throughout the entire space. The introduction of the constants \( C_1 \) and \( C_2 \) is necessary to avoid divergences in the scalar curvature and to ensure a smooth transition to the Euclidean regime at large distances. The conformal function \( \Omega(r) \) is given by,
\begin{equation}
\Omega(r) = -\frac{\lambda_0 r_0}{4\sqrt{\pi}} e^{-r^2 / r_0^2} \operatorname{erfc}(r/r_0) + C_1 \ln(r) + C_2.
\label{ConfFactor}
\end{equation}
which introduces an exponential decay modulated by the complementary error function \( \operatorname{erfc} \), ensuring controlled behavior both at the origin and at the asymptotic limit. The corresponding conformal metric is expressed as:
\begin{equation}
ds^2 = e^{-\frac{\lambda_0 r_0}{2\sqrt{\pi}} e^{-r^2 / r_0^2} \operatorname{erfc}(r/r_0) + 2C_1 \ln(r) + 2C_2} \left( dr^2 + r^2 d\theta^2 \right) + dz^2.
\end{equation}
This choice of conformal factor \( \Omega(r) \) ensures that the curvature remains finite and well-defined, allowing the defect to be treated in a regularized manner, avoiding the typical divergences of purely logarithmic models. Moreover, in the limit \( r \to \infty \), the conformal factor tends to a constant, which leads the metric to asymptotically approach the expected Euclidean form.

To guarantee that the metric is well-behaved and free of divergences, appropriate conditions must be imposed on the constants \( C_1 \) and \( C_2 \) presented in equation \eqref{ConfFactor}. It is evident that the term \( C_1 \ln(r) \) can introduce a singularity at \( r = 0 \). To avoid this, we must impose \( C_1 = 0 \) or choose a sufficiently small value to ensure that the metric remains regular at the origin. If \( C_1 \neq 0 \), it can dominate the metric at small \( r \), leading to a curvature divergence. Additionally, to ensure that the metric tends to the Euclidean form at large distances \( r \to \infty \), we must impose that \( e^{2\Omega(r)} \to 1 \), {\it i.e.}, \( \Omega(r) \to 0 \) or a finite constant. Since the first term in \( \Omega(r) \) vanishes as \( r \to \infty \), the choice of \( C_2 \) can be adjusted to guarantee this behavior. Thus, to prevent divergences, we impose \( C_1 = 0 \) and select \( C_2 \) such that \( e^{2C_2} = 1 \), which means that \( C_2 = 0 \) or another value that properly normalizes the conformal factor. By enforcing \( C_1 = 0 \) and selecting \( C_2 \) so that \( \Omega(r) \) remains finite at large distances, we ensure that the metric is well-defined throughout the entire space without introducing artificial singularities.
\begin{equation}
    \Omega(r) = -\frac{\lambda_0 r_0}{4\sqrt{\pi}} e^{-r^2 / r_0^2} \operatorname{erfc}(r/r_0).
\end{equation}
The complementary error function, denoted as \( \operatorname{erfc}(x) \), plays a fundamental role in describing the smooth transition between the central region of the defect and the asymptotically flat space. This function exhibits an exponential decay for large arguments, ensuring that the geometric perturbation induced by the defect remains localized. Mathematically, \( \operatorname{erfc}(x) \) is defined as:
\begin{equation}
    \operatorname{erfc}(x) = \frac{2}{\sqrt{\pi}} \int_x^\infty e^{-t^2} dt,
\end{equation}

\begin{figure}[H]
    \centering
    \includegraphics[width=0.6\textwidth]{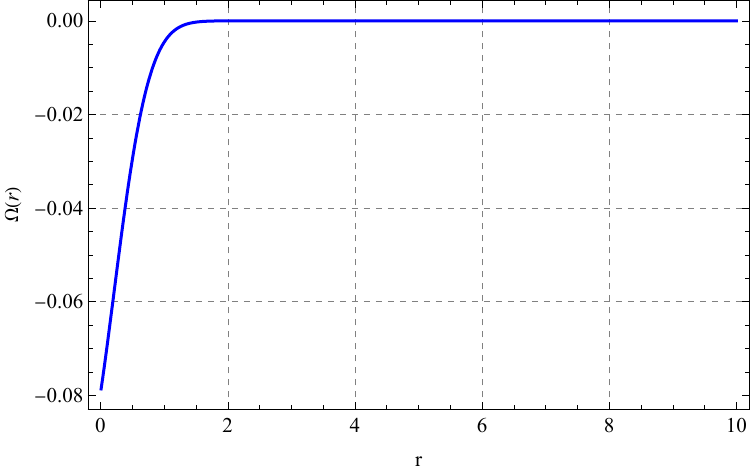}
    \caption{Conformal factor $\Omega(r)$ for a Gaussian distribution of topological defects. The function is regular and negative throughout the domain, decaying rapidly to zero as $r \to \infty$. This indicates that the geometric deformation is strongly localized near the origin, with negligible effects at larger distances.}
    \label{fig:omega_gaussian}
\end{figure}

The metric for the Gaussian distribution of defects can be written as
\begin{equation}
    ds^2 = e^{-\frac{\lambda_0 r_0}{2\sqrt{\pi}} e^{-r^2 / r_0^2} \operatorname{erfc}(r/r_0)} \left( dr^2 + r^2 d\theta^2 \right) + dz^2,
    \label{metricGauss1}
\end{equation}
and considering large distances, the conformal metric reduces to the following form
\begin{eqnarray}
    ds^2 \approx \left( 1 - \frac{\lambda_0 r_0}{2\pi} \frac{e^{-r^2 / r_0^2}}{r/r_0} \right) ( dr^2 + r^2 d\theta^2 ) + dz^2.
    \label{metricGauss2}
\end{eqnarray}
The metric describes a localized deformation of the space resulting from a Gaussian distribution of defects, with curvature concentrated near the defect. At large distances, the metric recovers the Euclidean form, ensuring that the perturbation remains confined. Unlike models based on Dirac delta distributions, this metric represents a smooth and finite deformation, without point singularities. Its application is relevant for modeling defects in two-dimensional materials, such as crystalline lattices and curved surfaces. As shown in Fig.~\ref{fig:omega_gaussian}, the conformal factor associated with a Gaussian distribution of defects exhibits a sharply localized behavior. The function $\Omega(r)$ starts from a finite negative value and rapidly decays to zero, reflecting the compactness of the deformation. This rapid decay contrasts with the slower saturation observed in the exponential case, indicating that the Gaussian geometry is more confined and induces weaker long-range effects.

\subsection{Topological analysis for a Gaussian defect distribution}

\begin{figure}[t!]
\centering
\begin{tabular}{cc}    
    \subfigure[
    Ricci scalar $R(r)$ for the Gaussian distribution of defects, shown in linear scale. The plot exhibits a bell-shaped curve (inverted) centered at the origin, indicating that the curvature is sharply localized around the defect core and vanishes rapidly as $r$ increases.        
        \label{fig:ricci_gaussian_linear}
    ]{
        \includegraphics[width=70mm]{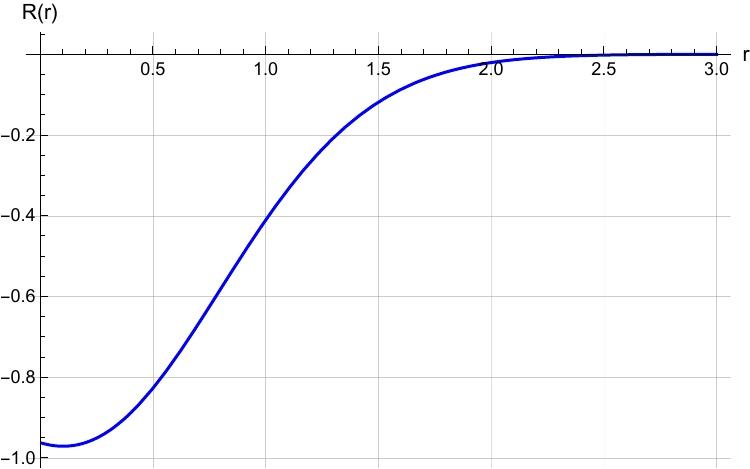}
    } &
    \subfigure[
        Logarithmic plot of the Ricci scalar magnitude $|R(r)|$ for the Gaussian distribution. The exponential decay is clearly visible, reinforcing the localized character of the curvature.
        \label{fig:ricci_gaussian_log}
    ]{
        \includegraphics[width=70mm]{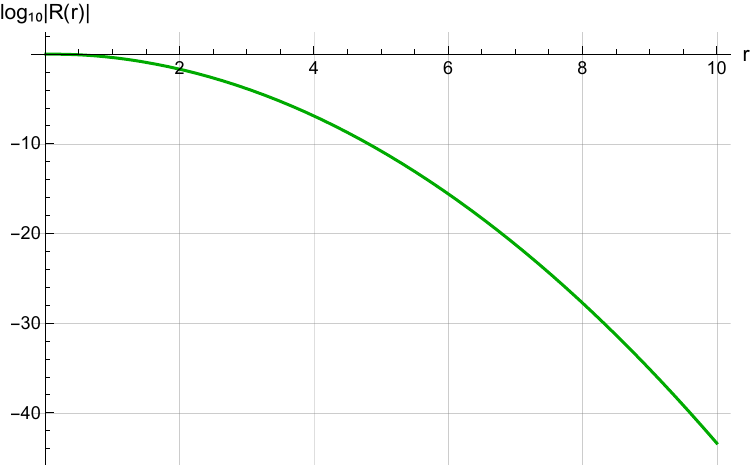}
    }
\end{tabular}
\caption{Ricci scalar plot for a Gaussian defect distribution}

\label{riccigauss}
\end{figure}

For now, let us analyze the Ricci scalar for a Gaussian distribution of defects given by the conformal metric \eqref{metricGauss1}. Through of relation $R = 2 e^{2\Omega} \Delta \Omega$, where we substitute \( \Omega(r) \) for our solution, we have that
\begin{eqnarray}
    R = -2 e^{2\Omega} \frac{\lambda_0}{\sqrt{\pi} r_0} e^{-r^2 / r_0^2}.
\end{eqnarray}
In the limit \( r_0 \to 0 \), the Ricci scalar reduces to:
\begin{eqnarray}
R \approx -2 e^{2\Omega} \lambda_0 \delta(r).
\end{eqnarray}
The set of figures \ref{riccigauss} shows the Ricci scalar related to the $r$ variable. As illustrated in Fig.~\ref{fig:ricci_gaussian_linear}, this plot confirms that the geometric deformation caused by a Gaussian distribution of defects is localized and concentrated near the center. The Ricci scalar has a clear minimum around $r = 0$ and quickly approaches zero, indicating the regularity of the solution and the topological neutrality at large distances. As demonstrated in Fig.~\ref{fig:ricci_gaussian_log}, the logarithmic scale emphasizes the exponential decay of the curvature, showing that geometric effects become negligible far from the defect. This behavior ensures that the total curvature remains finite and that the space effectively becomes Euclidean at infinity — a key condition for the topological consistency of the solution. This result aligns with the known expression for a localized topological defect modeled by a delta function, confirming that the Gaussian distribution functions as a smooth regularization of the singular defect. When $r_0 \to 0$, the smooth curvature reduces to a singularity at $r = 0$, reproducing the expected behavior of a point-like defect.

Return to the Gauss-Bonnet theorem for a surface without boundaries \eqref{Gauss-Bonet}, where \( K \) is the Gaussian curvature and \( \chi(S) \) is the Euler characteristic. The Gaussian curvature is given by
\begin{eqnarray}
    K = -\frac{\lambda_0}{\sqrt{\pi} r_0} e^{-r^2 / r_0^2}.
\end{eqnarray}
Computing the integral,
\begin{eqnarray}
    \int_S K dA = \int_0^{2\pi} d\theta \int_0^{\infty} K e^{2\Omega} r dr.
\end{eqnarray}
This result confirms that the contribution of curvature is finite and aligns with the Gauss-Bonnet theorem, ensuring topological consistency in our description of defect-induced geometry. The result of the total curvature integral can be written as
\begin{equation}
    \int_S K dA = -2\pi + 2\pi e^{-\frac{\lambda_0 r_0}{2\sqrt{\pi}}}.
    \label{Gauss-BonetResult}
\end{equation}
The outcome in equation~\eqref{Gauss-BonetResult} indicates that the total curvature is not exactly $-2\pi$, as in the case of point defects modeled by Dirac delta distributions, but rather a smoothed version due to the Gaussian defect distribution. The exponential in the second term of this result introduces a correction to the total curvature, which depends on the parameters $\lambda_0$ and $r_0$. This factor ensures that the curvature is not concentrated at a single point but rather distributed smoothly throughout the space. In other words, the smoothing effect prevents abrupt singularities and reflects a gradual transition in the geometry of space.

In the limit $\lambda_0 r_0 \to \infty$, the exponential term disappears, resulting in the total curvature of $-2\pi$, which corresponds to the classic case of a point-like defect. However, for finite values of $\lambda_0 r_0$, the total curvature is smaller in magnitude, emphasizing the regularization effect introduced by the Gaussian distribution. This behavior is expected in physical systems where curvature is not sharply concentrated but instead spreads out due to effects such as elasticity, microscopic interactions, or limits in the resolution of the defect's structure. A Gaussian defect distribution creates an even smoother curvature transition, ensuring that space changes occur gradually without sharp jumps. The finite and continuous topological charge in this scenario reflects the defect's distributed nature, with its influence confined spatially because of the rapid decay of the Gaussian function. Unlike the exponential model, where curvature effects extend over a larger area, the Gaussian distribution restricts the defect's impact to a more localized region. This behavior is especially important in superconductors and liquid crystals, where the size of topological defects affects phase transitions and transport processes.

\section{Defects with Curvature Density Inversely Proportional to Distance}

An alternative approach to modeling topological defects is to consider a curvature distribution that decays inversely with distance. For this case, the parameter $\lambda(r)$ can be expressed as
\begin{eqnarray}
    \lambda(r) = \frac{C}{r},
\end{eqnarray}
where $C$ is a constant associated with the intensity of the defect. This choice differs from traditional descriptions based on Dirac delta distributions, as it allows for a continuously distributed curvature field throughout space. Substituting this expression into the Poisson equation for the conformal factor, we obtain the solution as follows bellow,
\begin{eqnarray}
    \Omega(r) = C_1 \, \ln r + C_2,
\end{eqnarray}
where $C_1$ and $C_2$ are integration constants. Consequently, the metric takes the conformal form:
\begin{eqnarray}
    ds^2 = e^{2\Omega(r)} (dr^2 + r^2 d\theta^2) = e^{2C_2} r^{2C_1} (dr^2 + r^2 d\theta^2).
    \label{MetricInverse1}
\end{eqnarray}

In order to guarantee that the metric~\eqref{MetricInverse1} remains well-behaved at the origin, we require that the conformal factor does not introduce singularities beyond those inherent to the defect structure. Then, it is necessary to impose the following condition,
\begin{eqnarray}
    C_1 \geq 0,
\end{eqnarray}
which ensures that the metric remains finite and non-singular at $r = 0$, aside from the expected topological (conical) singularity localized at the core of the defect. Physically, the parameter $C_1$ controls the angular deficit associated with the curvature concentrated at the origin. In systems with discrete rotational symmetry, such as graphene, disclinations correspond to the insertion or removal of angular sectors of $\pi/3$, leading to a natural quantization of the angular deficit. This, in turn, constrains $C_1$ to take values in the discrete set $C_1 \in \mathbb{Z}/6$, according to the topology of the underlying lattice. And to ensure a physically meaningful asymptotic behavior, we also impose the boundary condition
\begin{eqnarray}
    \lim_{r \to \infty} \Omega(r) = 0,
\end{eqnarray}
which implies that $C_2 = 0$ up to an overall rescaling of the metric. This condition guarantees that the geometry approaches the flat Euclidean plane far from the defect.

To evaluate the curvature of this solution, we compute the Ricci scalar,
\begin{eqnarray}
    R = 0 \quad \text{for}\quad r > 0.
\end{eqnarray}
Thus, the space is locally flat everywhere except at the origin. However, the metric introduces a global topological modification, leading to a non-trivial angular deficit. Although $R = 0$, the Gauss-Bonnet theorem ensures that the total curvature, including contributions from the conical singularity, satisfies the following expression
\begin{eqnarray}
    \oint_{\partial S} k_g ds = 2\pi (1 - \alpha),
\end{eqnarray}
where the angular deficit is determined by
\begin{eqnarray}
    \alpha = \frac{1}{1+C_1}.
\end{eqnarray}
This shows that a nonzero angle deficit can exist even when the Ricci scalar vanishes. The effect is purely global and is associated with the topology of the space rather than with the local curvature.

A way to visualize this is through a change in variables, defining a new radial coordinate $\tilde{r} = r^{1+C_1}$, which reveals a modified angular structure. The total deficit angle is given by:
\begin{eqnarray}
    \delta = 2\pi (1 - \alpha) = 2\pi C_1.
\end{eqnarray}
Thus, even in a locally flat space, the topology is altered due to the conical singularity at the origin. This scenario is similar to disclinations in crystalline materials, where the curvature is localized at a single point, but the surrounding space remains flat. In such systems, the defect modifies the topology rather than the local curvature, leading to measurable effects on physical observables, such as transport properties in materials or geodesic deviations in spacetime. Although the curvature integral over the bulk is zero, the deficit angle contributes to the total curvature as captured by the Gauss-Bonnet theorem. The space retains a locally Euclidean structure but is globally altered.

When the defect density decreases inversely with distance, the induced curvature does not remain localized, but instead influences the geometry over a large region. The space exhibits a global angular deficit, which affects geodesic motion and alters how particles and fields behave in the presence of the defect. Unlike Gaussian and exponential distributions, where the effects diminish quickly, this type of decay modifies the structure of space in a more persistent way. This distribution is relevant in elasticity theory, where stress fields in materials can extend over long distances, as well as in analog gravity models, where defects introduce large-scale geometric distortions~\cite{nakahara2003geometry}.

\section{Defects with Curvature Density Inversely Proportional to the Square of the Distance}

Another relevant case arises when considering a curvature distribution that decays inversely with the square of the distance. In this scenario, the parameter $\lambda(r)$ is inversely proportional to the square of the radial variable and can be written as
\begin{eqnarray}
    \lambda(r) = \frac{\kappa}{r^2},
\end{eqnarray}
where $\kappa$ is a constant that characterizes the intensity of the defect. Substituting this expression into the Poisson equation for the conformal factor, we obtain the expression
\begin{eqnarray}
    \Omega(r) = (C_3 - \kappa) \ln r + C_4.
\end{eqnarray}
The corresponding conformal metric is given by
\begin{eqnarray}
    ds^2 = e^{2\Omega(r)} (dr^2 + r^2 d\theta^2) = e^{2C_4} r^{2(C_3 - \kappa)} (dr^2 + r^2 d\theta^2).
\end{eqnarray}
To determine $C_3$, we impose two conditions: 
\begin{enumerate}
    \item The regularity in origin, we see that the exponent $C_3 - \kappa$ must not introduce divergences at $r=0$. Therefore, we must impose \begin{eqnarray}
       C_3 - \kappa \geq 0.
    \end{eqnarray} 

    \item Asymptotic Flatness. The metric should approach the Euclidean form at infinity, so that
   \begin{eqnarray}
       \lim_{r \to \infty} \Omega(r) = 0.
   \end{eqnarray}
   This fixes $C_4 = 0$.
\end{enumerate}

The deficit angle associated with the topology of the space is given by:
\begin{eqnarray}
    \delta = 2\pi (C_3 - \kappa).
\end{eqnarray}
Thus, if a specific angular deficit is desired, we can choose $C_3$ accordingly:
\begin{eqnarray}
    C_3 = \kappa + \frac{\delta}{2\pi}.
\end{eqnarray}
This allows direct control over the global geometric properties of space through an appropriate choice of $C_3$. If no additional deficit is needed beyond what comes from the defect structure itself, setting $C_3 = \kappa$ ensures that the space remains asymptotically Euclidean while preserving regularity at the origin. By applying the conditions above, we guarantee a well-defined metric both locally and at large distances, eliminating ambiguities in choosing the constants of integration. A defect distribution that decreases quadratically with distance results in a geometry where the angular deficit is noticeable, yet the overall space remains nearly flat at large scales. The influence of the defect is confined to a finite region, making its effects highly localized compared to other distributions. This behavior is especially important in crystalline materials, where local curvature impacts mechanical and electronic properties. Additionally, this type of decay has implications in effective gravitational theories, where topological defects act similarly to mass distributions, affecting the behavior of fields and particles in lower-dimensional space-times\cite{carlip1998quantum,cai2010topological}.

\medskip

\noindent
\textbf{Remark on the general solution:} The general solution to the Poisson equation with defect densities of the form \( \lambda(r) \sim 1/r^n \) includes additional terms, such as \( -C r \) for \( n = 1 \) and \( -\frac{\kappa}{2} (\ln r)^2 \) for \( n = 2 \), which arise as particular solutions. However, in this work, we intentionally excluded these contributions because they could lead to physically undesirable behaviors. Specifically, the linear term \( -C r \) causes the conformal factor to decay exponentially at large distances, effectively confining the geometric deformation to a finite region around the defect. Consequently, the geometry becomes asymptotically flat, with the curvature vanishing at infinity. In contrast, the logarithmic squared term \( (\ln r)^2 \) produces strong divergences either at the origin or at spatial infinity, depending on its coefficient sign, resulting in unphysical behaviors in the curvature.

Since our aim is to build topologically nontrivial yet physically acceptable geometries - featuring finite curvature, controlled asymptotics, and a clear interpretation in terms of conical singularities - we only keep the homogeneous logarithmic contributions. These terms are directly linked to localized angular deficits and conical geometries, ensuring that the metric remains well-behaved and compatible with global geometric tools such as the Gauss-Bonnet theorem. We do not include plots for the cases $1/r$ and $1/r^2$, as their profiles are analytically simple and exhibit expected qualitative behavior: logarithmic or power-law divergence at large distances, without regularization at the origin.

\section{Conclusion}

In this study, we examined the geometric and topological effects of continuous distributions of topological defects in two-dimensional systems using a conformal metric framework. Unlike traditional models that rely on delta function singularities, we introduced smooth distributions, such as Gaussian, exponential, and power-law profiles, that regularize curvature while maintaining the essential topological features of the defects. Our analysis showed that Gaussian and exponential distributions produce regular, finite metrics with localized curvature near the defect core. The Gaussian profile causes the most compact deformation, with curvature rapidly diminishing and negligible effects at long distances. In contrast, the exponential distribution has a broader spatial influence, while still maintaining a finite curvature flux. These findings not only eliminate geometrical singularities but also offer a more realistic physical model of defects, relevant in condensed matter systems and emergent gravity analogs.

Power-law distributions, such as \( \lambda(r) \sim 1/r \) and \( \lambda(r) \sim 1/r^2 \), were also explored. They create conical geometries with angular deficits but almost zero Ricci curvature everywhere. Despite their local flatness, they induce global topological changes through nontrivial holonomy, making them excellent analogs for classical disclinations in crystalline materials. We confirmed that these geometries satisfy the Gauss-Bonnet theorem, verifying the consistency of the total curvature and surface topology. Overall, this research underscores the usefulness of smooth defect distributions in building physically consistent models that avoid singularities while capturing topological phenomena. These models could be applied to two-dimensional materials such as graphene, liquid crystals, and effective lower-dimensional gravity theories.

Future work includes studying quantum field propagation in these geometries, examining the role of torsion in models of screw dislocations, and exploring holonomy, wave scattering, and thermal transport in these curved backgrounds. Although not explicitly addressed here, an important next step is to analyze geodesic motion and holonomy in these settings, which can provide a deeper understanding of the effective geometry created by smooth defect distributions and further insight into the effective geometry induced by them.
{\bf Acknowledgments:} This work was supported by Conselho Nacional de Desenvolvimento Cient\'{\i}fico e Tecnol\'{o}gico (CNPq) and Funda\c{c}\~ao de Apoio a Pesquisa do Estado da Para\'iba (Fapesq-PB). G. Q. Garcia would like to thank Fapesq-PB for financial support (Grant BLD-ADT-A2377/2024). The work by C. Furtado is supported by the CNPq (project PQ Grant 1A No. 311781/2021-7).

\bibliographystyle{iopart-num}
  \bibliography{biblio} 

\providecommand{\newblock}{}
\begin{thebibliography}{10}
\expandafter\ifx\csname url\endcsname\relax
  \def\url#1{{\tt #1}}\fi
\expandafter\ifx\csname urlprefix\endcsname\relax\def\urlprefix{URL }\fi
\providecommand{\eprint}[2][]{\url{#2}}

\bibitem{Fumeron2021}
Fumeron S, Berche B and Moraes F 2021 {\em Liquid Crystals Reviews\/} {\bf 9} 85--110

\bibitem{Fumeron2023b}
Fumeron S and Berche B 2023 {\em The European Physical Journal Special Topics\/} {\bf 232} 1813--1833

\bibitem{mermin1979topological}
Mermin N~D 1979 {\em Reviews of Modern Physics\/} {\bf 51} 591--648

\bibitem{geroch1987strings}
Geroch R and Traschen J 1987 {\em Physical Review D\/} {\bf 36} 1017--1031

\bibitem{steinbauer2006use}
Steinbauer R and Vickers J~A 2006 {\em Classical and Quantum Gravity\/} {\bf 23} R91--R114

\bibitem{seung1988defects}
Seung H~S and Nelson D~R 1988 {\em Physical Review A\/} {\bf 38} 1005--1018

\bibitem{bowick2009two}
Bowick M~J and Giomi L 2009 {\em Advances in Physics\/} {\bf 58} 449--563

\bibitem{KatanaevVolovich1992}
Katanaev M~O and Volovich I~V 1992 {\em Annals of Physics\/} {\bf 216} 1--28

\bibitem{katanaev2005geometric}
Katanaev M~O 2005 {\em Physics-Uspekhi\/} {\bf 48} 675--701

\bibitem{DeserJackiw1984}
Deser S and Jackiw R 1984 {\em Annals of Physics\/} {\bf 153} 405--416

\bibitem{DeserJackiw1989}
Deser S and Jackiw R 1989 {\em Annals of Physics\/} {\bf 192} 352--367

\bibitem{deser1984three}
Deser S, Jackiw R and 't~Hooft G 1984 {\em Annals of Physics\/} {\bf 152} 220--235

\bibitem{carlip1998quantum}
Carlip S 1998 {\em Quantum Gravity in 2+1 Dimensions\/} (Cambridge University Press)

\bibitem{kleinert1989gauge}
Kleinert H 1989 {\em Gauge Fields in Condensed Matter, Vol. II: Stresses and Defects\/} (World Scientific)

\bibitem{nelson2002defects}
Nelson D~R 2002

\bibitem{zurek1996cosmological}
Zurek W~H 1996 {\em Physics Reports\/} {\bf 276} 177--221

\bibitem{Cortijo2007}
Cortijo A, Guinea F and Vozmediano M~A~H 2012 {\em Journal of Physics A: Mathematical and Theoretical\/} {\bf 45} 383001 originally posted to arXiv as 1112.2054v1 (\textit{Preprint} \eprint{1112.2054}) \urlprefix\url{https://arxiv.org/abs/1112.2054}

\bibitem{Eguchi1980}
Eguchi T, Gilkey P~B and Hanson A~J 1980 {\em Physics Reports\/} {\bf 66} 213--393

\bibitem{nakahara2003geometry}
Nakahara M 2003 {\em Geometry, Topology and Physics\/} (CRC Press)

\bibitem{chern1944gauss}
Chern S~S 1944 {\em Annals of Mathematics\/} {\bf 45} 747--752

\bibitem{hehl1976spin}
Hehl F~W, der Heyde P~V, Kerlick G~D and Nester J~M 1976 {\em Reviews of Modern Physics\/} {\bf 48} 393--416

\bibitem{Carvalho2013}
de~M~Carvalho A~M, de~Lima~Ribeiro C~A, Moraes F and Furtado C 2013 {\em The European Physical Journal Plus\/} {\bf 128} 60

\bibitem{Vitelli2004}
Vitelli V and Turner A~M 2004 {\em Phys. Rev. Lett.\/} {\bf 93} 215301 (\textit{Preprint} \eprint{cond-mat/0406329})

\bibitem{Schmitz2021}
Schmitz A~T, Spanton E~M and Grushin A~G 2021 {\em arXiv preprint\/} (\textit{Preprint} \eprint{2106.08520})

\bibitem{abramowitz1972handbook}
Abramowitz M and Stegun I~A 1972 {\em Handbook of Mathematical Functions with Formulas, Graphs, and Mathematical Tables\/} 10th ed (New York: Dover Publications) reprint of the 1964 edition published by the U.S. National Bureau of Standards

\bibitem{culetu2022modifiedrindler}
Culetu H 2022 {\em arXiv preprint\/} (\textit{Preprint} \eprint{2206.02550}) \urlprefix\url{https://arxiv.org/abs/2206.02550}

\bibitem{eguchi1980gravitation}
Eguchi T, Gilkey P~B and Hanson A~J 1980 {\em Physics Reports\/} {\bf 66} 213--393

\bibitem{Chavanis2009}
Chavanis P~H 2009 {\em The European Physical Journal B\/} {\bf 70} 413--433

\bibitem{cai2010topological}
Cai R~G, Cao L~M and Ohta N 2010 {\em Physical Review D\/} {\bf 81} 024003

\end{thebibliography}
\end{document}